\documentstyle[a4,12pt]{article}

\begin{document}

\newcommand{\zb}{\bar z}
\newcommand{\th}{\theta}
\newcommand{\tb}{\bar{\theta}}
\newcommand{\zt}{\tilde z}
\newcommand{\zbt}{\tilde{\zb}}
\newcommand{\tht}{\tilde{\th}}
\newcommand{\tbt}{\tilde{\tb}}
\newcommand{\TH}{\Theta}
\newcommand{\TB}{\bar{\Theta}}

\newcommand{\la}{\lambda}
\newcommand{\LA}{\Lambda}
\newcommand{\DE}{\Delta}
\newcommand{\OM}{\Omega}
\newcommand{\SS}{{\bf S\Sigma}}

\newcommand{\pa}{\partial}
\newcommand{\pab}{ \bar{\partial} }
\newcommand{\dab}{\bar D}

\newcommand{\pat}{ \tilde{\pa} }
\newcommand{\pabt}{ \tilde{\pab} }
\newcommand{\dt}{ \tilde{D} }
\newcommand{\dbt}{ \tilde{\dab} }

\newcommand{\ZB}{\bar Z}
\newcommand{\HT}{ {H_{\th}}^z }
\newcommand{\HB}{ {H_{\tb}}^z }
\newcommand{\HO}{ H_{\th} ^{\ \th} }
\newcommand{\HZ}{ H_{\zb} ^{\ \th} }
\newcommand{\HZB}{ H_{\zb} ^{\ z} }
\newcommand{\HOB}{ H_{\tb} ^{\ \th} }

\newcommand{\KT}{ {k_{\tb}}^{\zb} }
\newcommand{\KB}{ {k_{\th}}^{\zb} }
\newcommand{\KO}{ k_{\tb} ^{\ \tb} }
\newcommand{\KZ}{ k_{z} ^{\ \tb} }
\newcommand{\KOB}{ k_{\th} ^{\ \tb} }
\newcommand{\KZB}{ k _{z}^{\ \zb} }
\newcommand{\LB}{\bar{L}}
\newcommand{\BT}{\bar{T}}
\newcommand{\HKB}{ \hat{k} _{\th}^{\ \zb} }
\newcommand{\HKOB}{ \hat{k} _{\th}^{\ \tb} }
\newcommand{\HKZB}{ \hat{k} _{z}^{\ \zb} }
\newcommand{\HKZ}{ \hat{k} _{z}^{\ \tb} }

\newcommand{\ro}{ {\stackrel{\circ}{\rho}} }
\newcommand{\rb}{ \bar{\rho} }
\newcommand{\rob}{ {\stackrel{\circ}{\rb}} }
\newcommand{\RR}{ {\stackrel{\circ}{R}} }
\newcommand{\rr}{ {\stackrel{\circ}{r}} }
\newcommand{\cc}{ {\stackrel{\circ}{\chi}} }
\newcommand{\ee}{ {\stackrel{\circ}{\eta}} }
\newcommand{\GG}{ {\stackrel{\circ}{\Gamma}} }
\newcommand{\GAM}{ {\stackrel{\circ}{\gamma}}_{\TH} }
\newcommand{\gam}{ {\stackrel{\circ}{\gamma}}_{\th} }
\newcommand{\ga}{ {\stackrel{\circ}{\gamma}} }
\newcommand{\eb}{ \bar{\eta} }
\newcommand{\eeb}{ {\stackrel{\circ}{\eb}} }
\newcommand{\uuu}{ {\stackrel{\circ}{u}} }
\newcommand{\kk}{ {\stackrel{\circ}{\kappa}} }
\newcommand{\ttt}{ {\stackrel{\circ}{\tau}} }
\newcommand{\uu}{ {\stackrel{\circ}{\Upsilon}} }

\hfill MPI-Ph/92-38
\vskip 0.07truecm
\hfill LPTB 92-6

\thispagestyle{empty}

\bigskip
\begin{center}
{\bf \Huge{Relating Weyl and }}
\end{center}
\begin{center}
{\bf \Huge{diffeomorphism anomalies}}
\end{center}
\begin{center}
{\bf \Huge{on super Riemann surfaces}}
\end{center}
\bigskip
\bigskip
\centerline{\bf Jean-Pierre Ader$^{\, a \, \S}$, $\ $
Fran\c cois Gieres$^{\, b \, \dag}$, $\ $ Yves
Noirot$^{\, a}$}
\bigskip
\centerline{$^{a}${\it Laboratoire de Physique Th\'eorique$^{\, {\bf \ddag}}$}}
\centerline{\it Universit\'e de Bordeaux I}
\centerline{\it Rue du Solarium}
\centerline{\it F - 33170 - Gradignan}
\bigskip
\centerline{$^{b}${\it Max-Planck-Institut f\"ur Physik}}
\centerline{\it Werner-Heisenberg-Institut}
\centerline{\it F\"ohringer Ring 6}
\centerline{\it D - 8000 - M\"unchen 40}
\bigskip
\bigskip
\bigskip
\bigskip
\bigskip

\begin{abstract}

Starting from the Wess-Zumino action
associated to the super Weyl anomaly, we
determine the local counterterm
which allows to pass from this anomaly to the
chirally split
superdiffeomorphism anomaly
(as defined on a compact super Riemann surface without boundary).
The counterterm involves the graded extension of the Verlinde functional
and the results can be applied to the
study of holomorphic factorization
of partition functions
in superconformal field theory.

\end{abstract}
\bigskip

\nopagebreak
\begin{flushleft}
\rule{2in}{0.03cm} \\

{\footnotesize \ ${}^{\S}$
E-mail address: ADER@FRCPN11.}
\\  [-0.04cm]
{\footnotesize \ ${}^{\dag}$
Alexander von Humboldt Fellow.
E-mail address: FRG@DM0MPI11.}
\\  [-0.04cm]
{\footnotesize \ ${}^{\ddag}$
Unit\'e
Associ\'ee au CNRS, U.A. 764.} \\  [0.5cm]

MPI-Ph/92-38   \\
LPTB 92-6     \\
\vskip 0.07truecm
April 1992
\end{flushleft}

\newpage

\setcounter{page}{1}

\section{Introduction}

Quite recently, the authors of reference \cite{klt} have determined
the local counterterm to the effective action which allows
one to shift the Weyl anomaly to the factorized (i.e. chirally
split) diffeomorphism anomaly (as defined on an
arbitrary compact Riemann surface without boundary). This
result was then explored \cite{kls}
to prove
the existence and holomorphic factorization of partition
functions viewed as functionals of Beltrami coefficients
(the latter parametrizing the space of conformal
structures).
A natural question to ask is whether these results also
hold in superconformal field theory and in the present paper,
we address these issues and extend the analysis \cite{klt}
in a supersymmetric way.

Let us be more concrete and present an outline of the paper.
The arena we work on is a
compact super Riemann surface (SRS)
${\bf S\Sigma}$ without boundary\footnote{
Actually, we will only consider supercomplex structures
satisfying $\HT =0$ - see section 3 - because the chirally
split superdiffeomorphism anomaly is only known in this case
for higher genus surfaces \cite{dg1}.}
\cite{df}.
We start from the super Weyl anomaly,
${\cal A}_W [ \Omega ]$,
which depends on the supervielbein fields and on the
Weyl ghost $\Omega$ \cite{p} \cite{gmc}.
Furthermore, we consider
the chirally split form of
the superdiffeomorphism anomaly,
${\cal A}_D [ C^z ] + {\rm c.c.}$,
which depends on the super Beltrami coefficients and
on the ghosts $C^z, C^{\zb}$
parametrizing supercoordinate transformations
\cite{dg1} \cite{bol}.
The main goal
is to determine the local functional
$\Gamma_{loc}$ whose BRS variation
$s\Gamma_{loc}$ relates these anomalies (thus proving their
equivalence):
\begin{equation}
\label{0}
{\cal A}_W [ \Omega ] \ + \ s\, \Gamma_{loc} \ = \
{\cal A}_D [ C^z ] \ + \ {\rm c.c.}
\ \ \ .
\end{equation}
We start from the Wess-Zumino (WZ) action for the super Weyl anomaly
\cite{gmc} and successively construct the three local counterterms
contributing to $\Gamma_{loc}$.
The first and third of these terms represent the supersymmetric generalization
of the Liouville counterterm and of the Verlinde action \cite{v} \cite{klt},
respectively. The latter functional
admits various applications in conformal field
theory \cite{v} \cite{klt} \cite{kls} and plays a r\^ole
for the field-theoretical realization
of $W$-algebras \cite{dbg}.

While the construction of the super Liouville counterterm
does not pose any major problems,
the remainder of the calculations is technically very complicated.
For this reason, the corresponding details have been deferred
to appendix C.
Remarkably enough, the results take a compact and transparent form
after their projection to component fields in the WZ-gauge (section 9).
We conclude by commenting on the applications of
the results to superconformal field theory.

In appendices A and B, we elaborate on the approach \cite{dg}
to SRS's and superconformal models; the corresponding results
are applied in the main body of the text and are
of independent interest as a complement to the study \cite{dg}.
Although we rely on the latter work for the
parametrization of superconformal
structures, we would like to mention that other approaches
have previously been considered by various groups
\cite{gn}-\cite{bbg}.

As to our notation,
we label the local coordinates on ${\bf S \Sigma}$ by
$(z, \zb , \th, \tb )$. The canonical basis of the tangent and
cotangent spaces are, respectively, denoted by
\begin{equation}
\label{1}
D \ \equiv \ D_{\th} \ \equiv \  \pa_{\th} \, + \, \th \, \pa_z
\ \ \ \ \ \ , \ \ \ \ \ \
D^2 \ = \
\pa \ \equiv \ \pa_z
\ \ \ \ \ \ \ \ \
{\rm and \ \, c.c.}
\end{equation}
and by
\begin{equation}
e^z \ \equiv \ dz \, + \, \th \, d\th
\ \ \ \ \ \ \ , \ \ \ \ \ \ \ \
e^{\th} \ \equiv \  d\th
\ \ \ \ \ \ \ \ \ \  {\rm and \ \, c.c.}
\ \ \ .
\end{equation}
In conclusion, we recall that
the canonical 1-forms $e^z$ and $e^{\th}$ satisfy
the structure equations
\begin{equation}
d\, e^z \ = \ -\, e^{\th} \, e^{\th}
\ \ \ \ \
\ \ \ \ \ \ , \ \ \ \ \ \ \;
d\, e^{\th} \ = \ 0
\ \ \ .
\label{2}
\end{equation}

\section{Vielbein fields and super Weyl anomaly}

The basic variables of supergravity
are the supervielbein 1-forms $E^A$ \cite{wb}.
In a light-cone type basis in two dimensions, we have
$(E^A ) = (E^{++} , E^{--} , E^+ , E^- )$ which quantities
can be expressed in terms of super isothermal coordinates
$(Z, \ZB , \TH , \TB )$, e.g. for the spatial components
\begin{eqnarray}
\label{3}
E^{++} & = & e^Z \, \left[ \, \delta_Z ^{\, ++} \,
{\rm exp} ( \sigma )\,
\right]
\ \equiv \ e^Z \ \rho_Z
\\
E^{--} & = &
e^{\ZB} \, \left[ \, \delta_{\ZB} ^{\, --} \, {\rm exp} ( \sigma ) \,
\right]
\ \equiv \ e^{\ZB} \ \rho_{\ZB}
\ \ \ .
\nonumber
\end{eqnarray}
Here, $e^Z$ and $e^{\ZB}$ are the canonical 1-forms and
$\sigma$ is a scalar superfield \cite{ph}.
Thus, the ``supermetric"\footnote{For the definition and interpretation
of a supermetric, see the remarks in references \cite{fg}.} is given by
\[
ds^2 \ = \ E^{++} \, E^{--}  \ = \
e^Z \, e^{\ZB} \; \left( \, \rho_Z \, \rho_{\ZB} \, \right)
\ \equiv \
e^Z \, e^{\ZB} \   \rho_{Z\ZB}
\ \ \ .
\]
Under an infinitesimal super Weyl rescaling
with parameter $\OM$, the conformal factors
$\rho_{Z}$ and $\rho_{\ZB}$ transform according to
$\delta_{\OM} \rho_Z  =  \OM \rho_{Z} , \,
\delta_{\OM} \rho_{\ZB}  =  \OM \rho_{\ZB}$.

The only independent field strength of the theory is a scalar
superfield $S$ containing the
scalar curvature of
space-time
among its components. The field $S$ transforms as \cite{ph}
\[
\delta_{\OM} S \ = \ - \, \OM \, S \, - \, 2\,
\nabla_+ \nabla_- \OM
\ \ \ ,
\]
where $\nabla_+$ and $\nabla_-$ are the spinorial
components of the Lorentz-covariant derivative in superspace.

The {\em super Weyl anomaly}
${\cal A} _W$
and the associated {\em WZ
action }
$\Gamma_{W}$
have the form \cite{gmc}
\begin{eqnarray}
\label{6}
{\cal A} _W [ \OM, E ] & \equiv & K\, \int_{\SS}
d^4z \  E \; \OM \, S
\\
\Gamma_{W} [\phi , E ] & = &
K\, \int_{\SS}
d^4z \  E \, \left\{ \, (\nabla_+ \phi )\, ( \nabla _- \phi )
\; + \; S\, \phi \, \right\}
\nonumber  \\
\delta_{\OM} \Gamma_{W} [\phi , E ] & = &  -\,
{\cal A}_{W} [\OM, E]
\ \ \ .
\nonumber
\end{eqnarray}
Here, $K$ is a constant, $E \, \equiv \, {\rm sdet} \, (E_M ^{\ A})$
and $\phi$ is the Goldstone superfield transforming as
$\delta_{\OM} \phi = -\OM$.

If we express the functionals (\ref{6}) in terms of isothermal
coordinates, we obtain
\begin{eqnarray}
\label{7}
{\cal A} _{W} [ \OM , E ] & = & -K\, \int_{\SS}
d^4Z \ \, \OM \; D_{\TH} D_{\TB} \, {\rm ln} \, \rho_{Z\ZB}
\\
\Gamma _{W}  [\phi , E ] & = &
-K\, \int_{\SS}
d^4Z \   \left\{ \, \phi \, ( D_{\TH} D_{\TB} \phi )
\; + \;  \phi \,
 ( D_{\TH} D_{\TB} \, {\rm ln} \, \rho_{Z\ZB} ) \, \right\}
\ \ \ ,
\nonumber
\end{eqnarray}
where $D_{\TH} \equiv \pa_{\TH} + \TH \pa_Z$ and $D_{\TB} \equiv
\overline{D_{\TH}}$ belong to the canonical tangent space basis.

\section{Beltrami coefficients and superdiffeomorphism ano\-maly}

An atlas of superconformal coordinates $(Z, \ZB , \TH , \TB )$
on $\SS$ defines a supercomplex structure or, equivalently, a
superconformal class of vielbein fields. These structures are
parametrized by Beltrami coefficients $\HB, \HT$ (and c.c.),
see \cite{dg} and references therein.
The parametrization is described by choosing a reference
coordinate system, denoted by small coordinates
$(z, \zb , \th , \tb )$, and expressing the canonical 1-forms
of the coordinate system
$(Z, \ZB , \TH , \TB )$ with respect to the corresponding 1-forms
of the reference coordinate system \cite{dg}:
\begin{eqnarray}
e^Z & = & \left[ \, e^z \ + \ e^{\zb} \, \HZB \ + \ e^{\th} \, \HT \ + \
e^{\tb} \, \HB \, \right] \, \LA_z ^{\ Z}
\label{10}  \\
e^{\TH} & = &
\left[ \, e^z \ + \ e^{\zb} \, \HZB \ + \ e^{\th} \, \HT \ + \
e^{\tb} \, \HB \, \right] \, \tau_z ^{\ \TH} \ + \
\left[ \, e^{\th} \, \HO  \ + \ e^{\zb} \, \HZ \ + \
e^{\tb} \, \HOB \, \right] \, \sqrt{\LA_z ^{\ Z}}
\nonumber
\end{eqnarray}
(and c.c.).
By fiat,
the `$H$' are inert under super Weyl transformations.
The structure relations for $e^Z, e^{\ZB} ,e^{\TH} , e^{\TB} $ (
i.e. eqs.(\ref{2}) as written in terms of capital coordinates)
imply that all
the `$H$' depend only on two independent ones, namely $\HB, \HT$ (and
c.c.). Moreover, these relations imply that $\LA$ is an independent
factor
(satisfying a linear differential equation given below) and that $\tau$
depends on it and on the `$H$'.

As discussed in reference \cite{dg}, the restriction of the geometry
where $\HT =0$ (and c.c.) is compatible with superconformal
changes of coordinates, but it implies that the superdiffeomorphism
group has to be restricted to a subgroup leaving this equation
invariant. At the infinitesimal level, this restriction
is given by the relation
\begin{equation}
\label{11}
C^{\th} \ =\ \frac{1}{2} \, DC^z
\ \ \ \ \ \ \ \ {\rm and \ \, c.c.}
\ \ \ ,
\end{equation}
where
\begin{eqnarray}
\label{12}
C^z & \equiv & \Xi^z \ + \ \Xi^{\zb} \, H_{\zb} ^{\ z} \ + \
\Xi ^{\th} \, \HT \ + \ \Xi ^{\tb} \, \HB
\\
C^{\th} & \equiv  &
\Xi ^{\th} \, \HO \ + \ \Xi ^{\zb} \, \HZ \ + \
\Xi ^{\tb} \, H_{\tb} ^{\ \th}
\nonumber
\end{eqnarray}
parametrize superdiffeomorphisms generated by the vector field
\[
\Xi \cdot \pa \ \equiv \
\Xi^{z} (z , \zb , \th , \tb )\, \pa_z \ + \
\Xi^{\zb} (z , \zb , \th , \tb )\, \pa_{\zb} \ + \
\Xi^{\th} (z , \zb , \th , \tb )\, D_{\th} \ + \
\Xi^{\tb} (z , \zb , \th , \tb )\, D_{\tb}
\ \ \ .
\]

For convenience, we summarize the relations which hold
for $\HT = 0$ :
\begin{eqnarray}
\tau & = &  D \sqrt{\LA}
\ \ \ \ \ , \ \ \ \ \
\HO \ = \ 1
\ \ \ \ \ , \ \ \ \ \
{H}_{\tb} ^{\ \th} \ = \ - \frac{1}{2} \
D  \HB
\ \ \ \ \ \ \, {\rm and \ c.c.}
\label{14} \\
\HZB & = & ( \dab   -  \HB  \pa  ) \, \HB \, + \, (\HOB )^2
\ \ \  ,  \ \ \ \
\HZ \ = \ \frac{1}{2} \ D \HZB
\ \ \ \ \ \ \ \ \ {\rm and \ c.c.}
\ \ \ .
\nonumber
\end{eqnarray}
The integrating factor equations (IFEQ's) satisfied by the factor
$\LA$ then reduce to
\begin{eqnarray}
\label{15}
\left[ \, \dab \ - \ \HB \, \pa \ - \ \HOB \, D \, \right]
\, \LA & = & (\pa \HB ) \, \LA
\\
\left[ \, \pab \ - \ \HZB \, \pa \ - \  \HZ \, D \,
\right]
\, \LA & = & (\pa \HZB ) \, \LA
\ \ \ ,
\nonumber
\end{eqnarray}
the second equation being a consequence of the first one.

Under an infinitesimal superdiffeomorphism generated by the vector field
$\Xi \cdot \pa$, the coordinates
$(Z, \ZB , \TH , \TB )$ change according to
\begin{eqnarray}
s\TH & = & C^z \, \tau \; + \; C^{\th} \, \sqrt{\LA} \ \equiv \
{\cal C} ^{\TH}
\ \ \ \ \
\ \ \ \ \ \ \ \
\ \ \ \ \ \ \ \ {\rm and \ \, c.c.}
\label{17}
\\
sZ & = & C^z \, \LA \; - \; \TH \, (s\TH)  \ \equiv \
{\cal C} ^Z \; - \; \TH \,
s\TH
\ \ \ \ \ \ \ \ {\rm and \ \, c.c.}
\ \ \ .
\nonumber
\end{eqnarray}
Here and in the following, we assume that the `$\Xi$' and `$C$'
are ghost fields and that `$s$' is the BRS differential\footnote{The
$s$-operator is supposed to act from the right; the BRS-algebra
is graded by the ghost number, the Grassmann parity
being $s$-inert.}.
For $\HT=0$, the `$\Xi$' are restricted by the condition
(\ref{11}) and the induced variations of $\HB, \LA$ and
$C^z$ read
\begin{eqnarray}
s \HB & = &  \left[ \, \dab \ - \ \HB \, \pa
\ + \ \frac{1}{2} \, (D \HB ) D \, \right] \, C^z
\ + \  ( \pa \HB ) \, C^z
\label{18} \\
s\LA & = & \pa( C^z \,  \LA ) \ + \ \frac{1}{2} \, (DC^z ) \, D \LA
\nonumber \\
s C^z & = & -\, \left[ \, C^z \, \pa C^z \ +\ \frac{1}{4}\; (D C^z )^2
\, \right]
\ \ \ .
\nonumber
\end{eqnarray}
We note that the transformation law of $\LA$ can be rewritten
by virtue of the IFEQ (\ref{15}) as
\begin{equation}
\label{19}
s\LA \ = \ \Xi \cdot \pa  \, \LA   \ + \  N\, \LA
\end{equation}
where
\[
N \ \equiv \ \pa \Xi^z \; + \; (\pa \Xi^{\zb}) \, \HZB \; +  \;
( \pa \Xi^{\tb} ) \, \HB
\ \ \ .
\]

On a SRS with $\HT =0$, the {\em chirally split form
of the superdiffeomorphism anomaly} is given by \cite{dg1}
\begin{equation}
\label{20}
{\cal A}_D [C^z , \HB ]
\, + \, {\rm c.c.}
\ \equiv \  \frac{K}{2}
\int_{\SS} d^4 z \
C^z  \left[ \,  \pa ^2 D    +
3\, {\cal R} _{z\th} \, \pa    +    (D {\cal R}_{z\th} ) \, D    +    2\,
(\pa {\cal R}_{z \th} ) \, \right]    \HB
\, + \, {\rm c.c.}
\ ,
\end{equation}
where the superfield
${\cal R} _{z\th} (z, \th )$ is the component of a superprojective connection
\cite{bol}.
The WZ consistency conditions for the anomalies (\ref{7})
and (\ref{20}) read
\[
s\, {\cal A} _W [\OM , \rho , \HB ] \ = \ 0 \ = \
s\, {\cal A} _D [C^z  , \HB ]
\ \ \ .
\]
Here, the $s$-operation is defined by eqs.(\ref{17})-(\ref{19}) and by
\begin{eqnarray}
s\, \OM & = & \Xi \cdot \pa \; \OM
\nonumber  \\
s\, \rho_z & = &  \OM \, \rho_z \; + \; \Xi \cdot \pa \, \rho_z
\; + \; N \, \rho_z
\label{22}
\\
s\, \rho_{\zb} & = &  \OM \, \rho_{\zb} \; + \; \Xi \cdot \pa \,
\rho_{\zb} \; + \; \bar{N} \, \rho_{\zb}
\nonumber
\\
s\, {\cal R}_{z \th} & = & 0 \ =\
s\, \bar{{\cal R}} _{\zb \tb}
\nonumber
\end{eqnarray}
with $\Xi$ subject to eq.(\ref{11}) and $\rho_z, \rho_{\zb}$
related to $\rho_Z, \rho_{\ZB}$ by
\begin{equation}
\label{23}
\rho_z \ = \ \LA_z ^{\ Z} \; \rho_Z
\ \ \ \ \ \ \ , \ \ \ \ \ \ \
\rho_{\zb} \ = \ \bar{\LA} _{\zb} ^{\ \ZB}  \; \rho_{\ZB}
\ \ \ .
\end{equation}

Since we are dealing with generic SRS's (with $\HT = 0$),
a comment should be made concerning the transformation laws of the
basic variables
under superconformal changes of coordinates
$(z , \zb , \th , \tb ) \rightarrow
(z^{\prime} , \zb^{\prime} , \th^{\prime} , \tb^{\prime} )$:
these laws are given in terms of
${\rm e}^{-w} \equiv
D \th^{\prime}$ with $\dab w = 0$ and they read
\begin{eqnarray}
(\HB )^{\prime} & = &
{\rm e}^{\bar{w}} \;
{\rm e}^{-2w} \; \HB
\ \ \ \ \ \  , \ \ \ \ \ \
(C^z )^{\prime} \ = \
{\rm e}^{-2w} \; C^z
\nonumber
\\
(\rho_z )^{\prime} & = &
{\rm e}^{2w} \; \rho_z
\ \ \ \;
\ \ \ \ \ \ \ \ \ \ , \ \ \ \ \ \ \,
(\rho_{\zb} )^{\prime} \ = \
{\rm e}^{2\bar{w}} \; \rho_{\zb}
\label{23a}
\\
( {\cal R}_{z \th} )^{\prime} & = &
{\rm e}^{3w} \; [ \,
{\cal R}_{z \th}  \, - \,  {\cal S}(z^{\prime} , \th^{\prime} ;
z  , \th ) \, ]
\ \ \ ,
\nonumber
\end{eqnarray}
where ${\cal S}$ denotes the super Schwarzian derivative.

\section{Liouville counterterm}

The WZ-action for the super Weyl anomaly is invariant under
superdiffeomorphisms and therefore the last of eqs.(\ref{6}) can be
rewritten as
\begin{equation}
\label{25}
s \, \Gamma_{W} [\phi , E ] \ = \  -\,
{\cal A}_{W} [\OM, E]
\ \ \ ,
\end{equation}
where $s$ is defined by the equations given above and by
\begin{equation}
\label{26}
s \, \phi \ \equiv \  -\, \OM
\; + \; \Xi \cdot \pa \, \phi
\ \ \ .
\end{equation}
The superfunction $\phi$ can be expressed in terms of the superconformal
fields
$\rho_z, \rho_{\zb}$ introduced in eq.(\ref{23}) and some background
fields $\stackrel{\circ}{\rho} _z, \stackrel{\circ}{\rho} _{\zb}$
which are $s$-inert
($s\stackrel{\circ}{\rho} _z = 0 = s \stackrel{\circ}{\rho} _{\zb}$) :
\begin{equation}
\label{27}
2 \, \phi \ \equiv \  {\rm ln} \,
\stackrel{\circ}{\rho} _{z\zb} \; - \ {\rm ln} \,
\rho_{z\zb} \ =  \  {\rm ln} \,
\stackrel{\circ}{\rho} _{Z\ZB} \; - \ {\rm ln} \,
\rho_{Z\ZB}
\ \ \ .
\end{equation}
Here,
$\stackrel{\circ}{\rho} _{z\zb} \  \equiv  \
\stackrel{\circ}{\rho} _z \,
\stackrel{\circ}{\rho} _{\zb}, \
\rho_{z\zb} \equiv \rho_z \, \rho_{\zb}$
and
$\stackrel{\circ}{\rho} _z , \,
\stackrel{\circ}{\rho} _{\zb}$ are related to
$\stackrel{\circ}{\rho} _Z , \,
\stackrel{\circ}{\rho} _{\ZB}$ as in eq.(\ref{23}).

The {\em Liouville counterterm} is obtained by substituting
eq.(\ref{27}) into the expression (\ref{7}) for the WZ-action :
\begin{eqnarray}
\Gamma_{Liouville} [ \rho, H ; \ro ] & \equiv &
\Gamma_{W} [ \phi, E ]
\\
& = & \frac{K}{4} \, \int_{\SS} d^4 Z \ \left\{ \,
\left( \, {\rm ln} \, \rho_{Z\ZB} \, - \, {\rm ln} \, \ro_{Z\ZB}
\, \right) \, D_{\TH} D_{\TB} \,
\left( \, {\rm ln} \, \rho_{Z\ZB} \, - \, {\rm ln} \, \ro_{Z\ZB}
\, \right) \right.
\nonumber
\\
& & \left. \ \ \ \ \ \ \ \ \ \ \ \ \  + \; 2\,
\left( \, {\rm ln} \, \rho_{Z\ZB} \, - \, {\rm ln} \, \ro_{Z\ZB}
\, \right) \, D_{\TH} D_{\TB} \,
\left( \, {\rm ln} \, \ro_{Z\ZB} \, \right) \, \right\}
\ \ \ .
\nonumber
\end{eqnarray}
To evaluate the response of this functional
to the $s$-variation
induced by the transformations (\ref{17}),
we use the method outlined in appendix A of reference \cite{dg1}
from which it follows that
\begin{eqnarray*}
s\, (d^4 Z ) & = & d^4 Z \,
\left[ \, \pa_Z {\cal C}^Z \, - \, D_{\TH} {\cal C}^{\TH}
\, + \, {\rm c.c.} \, \right]
\\
s\, \pa_Z & = &   - \,
\left[ \, (\pa_Z {\cal C}^Z ) \pa_Z \, + \, (\pa_Z {\cal C}^{\ZB} )
\pa_{\ZB} \, + \,
(\pa_Z {\cal C}^{\TH} ) D_{\TH} \, + \,
(\pa_Z {\cal C}^{\TB} ) D_{\TB} \, \right]
\ \ \ \ \ \ \ \ \ \ \ \ \ \,
\ \ \ \ \ {\rm and \ c.c.}
\\
s\, D_{\TH} & = &   - \,
\left[ \, (D_{\TH} {\cal C}^Z ) \pa_Z \, - \, 2\,  {\cal C}^{\TH} \pa_Z
\, + \, (D_{\TH} {\cal C}^{\ZB} )
\pa_{\ZB} \, + \,
(D_{\TH}  {\cal C}^{\TH} ) D_{\TH} \, + \,
(D_{\TH} {\cal C}^{\TB} ) D_{\TB} \, \right]
\ {\rm and \ c.c.}
\ .
\end{eqnarray*}
Here,
${\cal C}^Z$ and ${\cal C}^{\TH}$ are the parameters introduced in
eqs.(\ref{17}).
By virtue of the previous relations and
\begin{equation}
\label{31}
{\cal C}^Z \pa_Z \, + \, {\cal C}^{\TH} D_{\TH} \, + \, {\rm c.c.}
\ = \ \Xi \cdot \pa
\ \ \ ,
\end{equation}
we find that
\begin{equation}
\label{30}
s\, \Gamma_{Liouville} \ = \
-\, {\cal A}_W [\OM, \rho , H] \ + \  \frac{K}{2} \,
\int_{\SS} d^4 Z \,
\left[ \, \Xi \cdot \pa \, {\rm ln} \; \ro_{z\zb}    +    N    +
\bar{N} \, \right]
D_{\TH} D_{\TB} \,
{\rm ln} \, \ro_{Z\ZB}
\ .
\end{equation}
The first term on the r.h.s. is the super Weyl anomaly
and the second represents
the {\em non-chirally split form of the superdiffeomorphism
anomaly} :
this term corresponds
\cite{bbs} \cite{klt} to a Weyl anomaly with $\rho$ replaced by
$\ro$ and $\OM$ replaced by the compensating value $\OM_{comp}$
which is obtained by setting $\rho = \ro$ in
the $s$-variation of ${\rm ln} \, \rho$ (see eqs.(\ref{22})):
\[
\Omega_{comp} \ = \ -\frac{1}{2} \, \left[ \, \Xi \cdot \pa \, {\rm ln}
\, \ro_{z\zb} \, + \, N \, + \, \bar{N}  \, \right]
\ \ \ .
\]

\section{Passage to a chirally split expression}

The next (and most difficult) step consists of establishing
a link between the integral on the r.h.s. of eq.(\ref{30}) and a
holomorphically factorized expression.
This process involves the passage from the capital to the
small coordinates and is technically very complicated.
Therefore, we will only summarize the results and defer all
details of the derivation to appendix C.
The main ingredient of the calculation are the so-called
`intermediate' or `tilde' coordinates
$(\tilde{z} , \tilde{\zb} , \tilde{\th} , \tilde{\tb} )
\equiv  ( Z , \zb , \TH, \tb )$
which mediate between the small and capital coordinates \cite{dg}.
The relevant equations
are presented in appendix A and here we only note that these coordinates
are {\em not} related to each other by
complex conjugation.

By virtue of eqs.(\ref{23}), (\ref{31}), (\ref{15}) and
(\ref{12}), the integral in eq.(\ref{30}) can be rewritten as
\begin{eqnarray}
\label{32}
\int_{{\bf S\Sigma}}
d^4 Z
\left[    \Xi \cdot \pa  {\rm ln}  \,\ro_{z\zb}    +    N    +
\bar{N}    \right]
D_{\TH} D_{\TB}
{\rm ln} \, \ro_{Z\ZB}
& = &
\int_{{\bf S\Sigma}}
d^4 Z
\left\{
\left[    {\cal C}^Z
\pa_Z + {\cal C}^{\TH} D_{\TH}
+ {\rm c.c.} \right]
{\rm ln} \, \ro_{Z\ZB}
\right.
\\
\nonumber
& &   +
\left.
\left[    \frac{1}{\LA} ( \pa
{\cal C}^Z ) + C^{\th} (D
{\rm ln} \, \LA )
+ {\rm c.c.} \right]  \right\}
D_{\TH} D_{\TB}
{\rm ln} \, \ro_{Z\ZB}
 .
\end{eqnarray}
The variables $C^{\th}$ and ${\cal C}^{\TH}$ depend on
the derivatives of $C^z$ according to eq.(\ref{11})
from which it follows that
\begin{eqnarray}
{\cal C}^{\TH} & = & \frac{1}{2 \sqrt{\LA}} \ D {\cal C}^Z
\ = \
\frac{1}{2} \ \tilde{D} {\cal C}^Z
\nonumber   \\
& = &
\frac{1}{2} \ \left[ \, D_{\TH}
+ ( \KB \LB ) \pa_{\ZB}
+ ( \KB \BT  + \KOB \sqrt{\LB} ) D_{\TB}
\right] {\cal C}^Z
\ \ \ .
\label{33}
\end{eqnarray}
Here, the coefficients `$k$' and the factors $\LB, \BT$ describe the
passage from the tilde to the capital coordinates, see appendix A.
As outlined in appendix C, the last equation and integration
by parts allow us to recast the integral (\ref{32}) in the form
\begin{eqnarray}
 & &
\int_{{\bf S\Sigma}}
d^4 Z \;
{\cal C}^Z  D_{\TB}
\left\{
\left[    \pat +
\frac{\pat \KT}{(\KO )^2}
(\dbt - \KT \pabt )
\right] \GAM
\right.
\label{34}
\\
 & &
\left.
\ \ \ \ \ \ \ \ \ \ \ \ \ \ \
\ \ \ \ \
\ \ \ \ \
- \, \frac{1}{2} \GAM
\left[    \dt +
\frac{\dt \KT}{(\KO )^2}
(\dbt - \KT \pabt )
\right] \GAM
\, - \, \frac{1}{2} (\KB \LB ) (D_{\TB}
\GAM )^2 \right\}
+ {\rm c.c.}
,
\nonumber
\end{eqnarray}
where
\begin{equation}
\label{35}
\GAM \ \equiv \ D_{\TH} \, {\rm ln} \, \ro_{Z \ZB}
\ \ \ .
\end{equation}
{}From (\ref{23}), (\ref{15}) and the explicit form of
the `$k$', one concludes that the expression
(\ref{35}) is related to the background connection
$\gam \equiv D \, {\rm ln} \, \ro_{z \zb}$ by
\begin{equation}
\label{36}
\GAM \ = \ \frac{1}{\sqrt{\LA}} \ \left( G \, - \,
D \, {\rm ln} \, \LA \, \right)
\ \ \ ,
\end{equation}
where $G$ equals
$\gam$ plus a contribution which does not depend on the integrating
factors $\LA , \bar{\LA}$.
Explicitly,
\begin{eqnarray}
\label{37}
G & = & \gam \, - \,
\bar{\nabla} _{\ga} ^2 H_{\th}^{\ \zb}
\, + \, ( \sqrt{\LA} \, \HKOB ) \, \left[
\nabla_{\ga} ^2 \HB  \, - \, \HB \,
(\bar{\nabla} _{\ga} ^2 H_z^{\ \zb} ) \, - \, \HOB \,
( \bar{\nabla} _{\ga} ^2 H_{\th}^{\ \zb} )
\right]
\\
& & \ \ \ \
\ \ \
\ \ \ \ \
\ \ \ \ \ \
+ \; ( \sqrt{\LA} \, \HKB ) \, \left[
\nabla_{\ga} ^2 \HZB \, - \, \HZB \,
(\bar{\nabla} _{\ga} ^2 H_z^{\ \zb} ) \, - \, \HZ \,
( \bar{\nabla} _{\ga} ^2 H_{\th}^{\ \zb} )
\right]
\nonumber
\end{eqnarray}
with $
\HKOB, \HKB$ given by eqs.(\ref{a3}) and $\nabla$ being the
{\em superconformally covariant derivative} introduced in
reference \cite{bol}:
\begin{equation}
\label{38}
\nabla_{\ga} F_p
\  \equiv \ ( \, D  -  \frac{1}{2} \, p \, \gam \, ) \, F_p
\ \ \ \ \ \ \ , \ \ \ \ \ \
\nabla_{\ga}^2 F_p
\ \equiv \
\nabla_{\ga} (
\nabla_{\ga} F_p )
\nonumber
\end{equation}
($F_p \ = \ $ conformal superfield of weight $p$ in the
($z, \th$)-sector, e.g. $C^z$ has weight `-2').
In fact, eqs.(\ref{23a}) imply that $\gam$ transforms like a
(non-holomorphic)
super affine connection under a superconformal change
of coordinates:
\begin{equation}
\label{666}
\left( \gam \right) ^{\prime} \ = \ {\rm e}^w \,
\left[ \, \gam \, + \, 2 \, (D w) \, \right]
\ \ \ .
\end{equation}

Further use of the relations in
appendix A yields the following
expression for the integral (\ref{34}):
\begin{equation}
\label{39}
\int_{{\bf S\Sigma}}
d^4 z \
\frac{C^z}{\KO}
\left\{
\nabla_{G} ^2 \psi     -
\left( \frac{\pa \KO}{\KO} - \frac{1}{2} G
\frac{D \KO}{\KO} \right) \psi    +
\sqrt{\LA} \KB \psi \left[ \pab G -
\nabla_{G} ^3 \HZB \right]  +
\sqrt{\LA} \KOB \psi^2
\right\}
+ {\rm c.c.}
{}.
\end{equation}
Here, $
\nabla_{G}$ denotes the superconformally
covariant derivative w.r.t. $G$
and the quantity
\begin{eqnarray}
\label{40}
\psi & \equiv &
\frac{1}{\KO}
\left\{
\left[ \dab G +
\nabla_{G} ^3 \HB \right]  - \KT
\left[ \pab G -
\nabla_{G} ^3 \HZB \right] \right\}
\\
& = &
\frac{1}{\KO}
\left\{
\left[ \pa D \HB + \KT \pa D \HZB \right] - \frac{1}{2}
\left[ \pa  \HB - \KT \pa  \HZB \right] G
+ ( \dbt - \KT \pabt ) G
\right\}
\nonumber
\end{eqnarray}
is independent of the integrating factors
$\LA,  \bar{\LA}$.
Thereby, the integral (\ref{39}) also has the property
that it does not depend on
$\LA$ and $\bar{\LA}$; this is a crucial point, since these factors are
non-local functionals
of $\HB$ by virtue of the IFEQ's (\ref{15}).

To summarize the derivation of this section, we have shown that the
second term on the r.h.s. of eq.(\ref{30}) has a chirally split form
in the sense that it represents
a linear functional of $C^z$ plus its
complex conjugate. In the next section, this expression will be related
to the chirally split superdiffeomorphism anomaly.

\section{Passage to the chirally split superdiffeomorphism
ano\-maly: the second counterterm}

By substituting eqs.(\ref{37}) and (\ref{40}) into (\ref{39}), we
end up with the following expression for the integral on the r.h.s.
of eq.(\ref{30}):
\begin{equation}
\label{42}
K \,
\int_{{\bf S\Sigma}}
d^4 z  \  C^z  \, \left\{ \,
\frac{1}{2} \, \pa^2 D \HB + \left[ \dab - \HB \pa +
\frac{1}{2} (D\HB) D -
\frac{3}{2} (\pa \HB) \right] \RR  \,
\right\}
\ + \  {\rm c.c.} \ + \ M
\ \ \ .
\nonumber
\end{equation}
Here,
\begin{equation}
\label{42a}
\RR \ \equiv \ \RR_{z\th} \ \equiv \
\frac{1}{2} \left[ \, \pa \gam \, - \, \frac{1}{2} \,
\gam \, (D\gam ) \, \right]
\end{equation}
represents a smooth superprojective connection
\cite{bol} and we have only specified contributions coming from
the leading term $\gam$ of $G$; all other contributions are included
in the quantity $M$, the latter being
lengthy and not very illuminating, cf. appendix B.

Let us now consider the supersymmetric generalization
of the {\em second counterterm} introduced in reference \cite{klt}:
\begin{equation}
\label{101a}
\Gamma_{II} \ \equiv \
- K \, \int_{{\bf S\Sigma}} d^4z \ \HB \, ( {\cal R} - \RR )
\ + \ {\rm c.c.}
 \ \ \ .
\end{equation}
Here, ${\cal R} \equiv
{\cal R}_{z\th}$ and $\RR \equiv \RR_{z\th}$ are
holomorphic and smooth superprojective
connections, respectively,
as introduced in eqs.(\ref{20}) and (\ref{42a}).

{}From eqs.(\ref{18})(\ref{20}) and the $s$-invariance of
${\cal R}$ and $\ro_{z\zb}$, it follows that
\begin{equation}
s \Gamma_{II} \; = \;     {\cal A}_D  \;  - \;
K
\int_{{\bf S\Sigma}}
d^4 z   \, C^z
\left\{
\frac{1}{2} \pa^2 D \HB + \left[ \dab - \HB \pa +
\frac{1}{2} (D\HB) D -
\frac{3}{2} (\pa \HB) \right] \RR
\right\}
\; +  \; {\rm c.c.}
\, .
\label{102}
\end{equation}
Obviously,
the integrals in expressions (\ref{42}) and
(\ref{102})
coincide with each other:
combining eqs.(\ref{30})(\ref{42}) and (\ref{102}),
we  get
\begin{equation}
\label{400}
{\cal A}_W [ \Omega , E ] \, + \, s\, \left(
\Gamma_{Liouville} + \Gamma_{II}  \right)
\ = \
{\cal A}_D [ C^z , \HB ] \, + \, {\rm  c.c.} \, + \, M
\ \ \ .
\end{equation}

\section{Verlinde's functional and the third counterterm}

The next step consists of determining the supersymmetric generalization
of the Verlinde functional \cite{v} on the superplane
and then turning
it into a globally well-defined
action on a SRS by the inclusion of connection terms
\cite{klt}. The final point (to be considered in the next section)
is to check that the $s$-variation of
this functional coincides with the quantity $M$ on the r.h.s.
of eq.(\ref{400}).

To supersymmetrize the
bosonic Verlinde action as defined on the complex plane ${\bf C}$,
we express the latter
in terms of the capital coordinates of the bosonic theory\footnote{
The latter are given in terms of reference coordinates ($z, \zb $)
by $dZ \, = \, \la (z, \zb ) \,
\left[ \, dz \, + \, \mu (z , \zb ) \, d\zb \, \right]$.} :
\begin{eqnarray*}
\Gamma_{Verlinde} \, [ \mu , \bar{\mu} ] & \equiv &
\int_{{\bf C}}
d^2 z \ \frac{1}{1 - \mu \bar{\mu}} \ \left\{ \,
(\pa \mu ) \, ( \pab \bar{\mu} ) \; - \; \frac{1}{2} \, \left[ \,
\bar{\mu} \, (\pa \mu )^2 \, + \, {\rm c.c.} \, \right] \, \right\}
\\
& = &
\int_{{\bf C}}
d^2 Z \ \left\{ \,
(\pa_{\ZB} \, {\rm ln} \, \la  ) \, ( \pa_Z \, {\rm ln} \, \bar{\la})
\; - \; \frac{1}{2} \, \left[
\, \bar{\mu} \; \frac{\bar{\la}}{\la} \;
(\pa_{\ZB} \, {\rm ln} \, \la  )^2
\, + \, {\rm c.c.} \, \right] \, \right\}
\ \ \ .
\end{eqnarray*}
Using this trick, we can immediately write down the
graded generalization of the last expression
on the superplane ${\bf SC}$:
\begin{equation}
\Gamma_{Verlinde} ^{(super)} [ \HB , H_{\th} ^{\ \zb} ]
 =
\int_{{\bf SC}}
d^4 Z  \left\{
(D_{\TB} \, {\rm ln} \, \LA )
(D_{\TH} \, {\rm ln} \, \bar{\LA} )
+ \frac{1}{2}  \left[
H_{\th} ^{\ \zb} \frac{\bar{\LA}}{\sqrt{\LA}} \,
(D_{\TB} \, {\rm ln} \, \LA )  ( \pa_{\ZB} \, {\rm ln} \, \LA \, )
 +  {\rm c.c.}
\right] \right\}
 .
\label{600}
\end{equation}
This functional
can easily be expressed in terms of the small coordinates by virtue of
eqs.(\ref{a2}) and the IFEQ's (\ref{a1b})(\ref{a9}):
\begin{eqnarray}
\Gamma_{Verlinde} ^{(super)}
& = &
\int_{{\bf SC}}
d^4 z \
\frac{1}{(\KO)^2}
\left\{ \,
[ \pa \HB - \KT (\pa \HZB ) ]
\left[ \,
[ \, 1 +
( \sqrt{\LA} \, \HKOB ) \HOB +
( \sqrt{\LA} \, \HKB ) \HZ \, ] \, \pab H_{\th} ^{\ \zb}
\right. \right.
\nonumber
\\
& &  \left. \left.
\ \ \ \ \ \ \ \ \ \ \ \ \ \ \
\ \ \ \ \ \ \ \ \ \ \ \ \ \ \ \ \
\ \ \ \ \ \ \ \ \ \ \ \ \ \ \ \ \
+  \; [ \,
( \sqrt{\LA} \, \HKOB ) \HB +
( \sqrt{\LA} \, \HKB ) \HZB \, ] \, \pab H_{z} ^{\ \zb} \,
\right]
\right.
\nonumber
\\
& &  \left.
\ \ \ \ \ \ \ \ \ \
\ \ \ \ \ \ \ \ \ \ \ \ \ \ \ \ \
+ \  \frac{1}{2} \, \left[ \,
\frac{1}{A} \, H_{\th} ^{\ \zb} \,
[ \pa \HB - \KT (\pa \HZB ) \, ] \,
\pa \HZB \, + \, {\rm c.c.} \, \right] \, \right\}
\ \ .
\label{601}
\end{eqnarray}
Here, the quantity $A$ is a function of the Beltrami coefficients
(explicitly given by eq.(\ref{157}) of appendix A).
Although most of the `$H$' and `$k$' transform in a complicated way under
superconformal transformations, it can be checked that the {\em super
Verlinde action} (\ref{601}) becomes
globally well-defined on ${\bf S \Sigma}$,
if the derivatives appearing explicitly in this expression
are replaced
by supercovariant ones (see eqs.(\ref{a21}).
This procedure provides the {\em third counterterm}
$\Gamma_{III} \, [ \HB , H_{\th} ^{\ \zb} ; \ro ]$:
\begin{eqnarray}
\Gamma_{III}
& = &
-K \int_{{\bf S \Sigma}}
d^4 z \,
\frac{1}{(\KO)^2} \,
\left\{
[ \nabla_{\ga}^2 \HB - \KT (\nabla_{\ga}^2 \HZB ) ]
\left[
[ 1 +
( \sqrt{\LA} \HKOB ) \HOB +
( \sqrt{\LA} \HKB ) \HZ ] \bar{\nabla} _{\ga}^2 H_{\th} ^{\ \zb}
\right. \right.
\nonumber
\\
& &  \left. \left.
\ \ \ \ \ \ \ \ \ \ \ \ \ \ \ \ \ \ \ \ \
\ \ \ \ \ \ \ \ \ \ \ \ \ \ \ \ \
\ \ \ \ \ \ \ \ \ \ \ \ \ \ \ \ \ \ \ \
+  \; [ \,
( \sqrt{\LA} \, \HKOB ) \HB +
( \sqrt{\LA} \, \HKB ) \HZB \, ] \, \bar{\nabla} _{\ga}^2 H_{z} ^{\ \zb}
\, \right]
\right.
\nonumber
\\
& &  \left.
\ \ \ \ \ \ \ \ \ \ \ \
\ \ \ \ \ \ \ \ \ \ \ \ \ \ \ \ \
+ \ \frac{1}{2} \, \left[ \,
\frac{1}{A} \, H_{\th} ^{\ \zb} \,
[ \, \nabla_{\ga}^2 \HB - \KT (\nabla_{\ga}^2 \HZB ) \, ] \,
\nabla_{\ga}^2 \HZB \, + \, {\rm c.c.} \, \right] \, \right\}
\ \ .
\label{602}
\end{eqnarray}

\section{Synthesis}

It remains to show that $s \Gamma_{III}$ coincides up to a sign
with the quantity $M$ occurring on the
r.h.s. of eq.(\ref{400}).
Since the $s$-variation of $\HB$ (and its c.c.) are known,
there is in principle no obstruction for determining the $s$-variations of the
`$k$' and of $\Gamma_{III}$. However, the latter expressions
represent very complicated functions of $\HB$ and $H_{\th} ^{\ \zb}$
and therefore a general solution of this problem
within a reasonable amount of time
appears to be out of reach.
For this reason, we rather solve this problem
in the WZ-gauge. The corresponding equations are explicitly
given below and our final result reads:
\begin{equation}
\label{500}
{\cal A}_W [ \Omega , E ] \, + \, s\, \left(
\Gamma_{Liouville} + \Gamma_{II}  + \Gamma_{III}
\right)
\ = \
{\cal A}_D [ C^z , \HB ] \, + \, {\rm  c.c.}
\ \ \ .
\end{equation}
It is unlikely that the validity of this result is destroyed
by the inclusion of auxiliary fields which are absent
in the WZ-gauge; in any case, the only possible modification consists of
a term vanishing in the WZ-gauge (see also appendix B).

\section{Projection to component fields}

The super Weyl anomaly and the associated WZ action,
as given by eqs.(\ref{6}),
are superspace integrals
involving the supervielbein fields: the corresponding component field
expressions immediately follow by application
of the so-called density projection
formula \cite{gn}.

The holomorphic superfield ${\cal R}$ admits a $\th$-expansion
of the form \cite{bol}
\begin{equation}
\label {300}
{\cal R}_{z\th} \ = \ \frac{i}{2} \, \chi_{z\th}
\, + \, \th
\, [ \frac{1}{2} \, r_{zz} ]
\ \ \ ,
\end{equation}
where $\chi$ and $r$ only depend on $z$ and not on $\zb$.
All other component fields to appear in this section depend on both
$z$ and $\zb$.

In the WZ-supergauge, we have \cite{dg}
\begin{equation}
\label {301}
\HB \ = \
\tb \, \mu_{\zb} ^{\ z} \, + \, \th \tb \, [ -i \alpha_{\zb} ^ {\ \th}  ]
\ \ \ \ \ \ , \ \ \ \ \ \ \
C^z \ = \  c^z \, + \, \th \, [ i \epsilon^{\th} ]
\ \ \ .
\end{equation}
Here,
the space-time fields $\mu$ and $\alpha$ denote the Beltrami coefficient and
its
fermionic partner, respectively, while $c$ and $\epsilon$ parametrize
ordinary diffeomorphisms and local supersymmetry transformations, respectively.
In the following, we will simplify the notation by suppressing all indices
on the component fields.

The $\th$-expansions of the (real) supermetric $\ro_{z\zb}$ and
of the superaffine connection $\gam$ read
\begin{eqnarray}
\ro_{z\zb} & = & \ro_0 \, + \, \th \, [i \ro_1] \, + \,
\tb \, [-i \rob_1 ] \, + \, \th \tb \, [ i \ro_2]
\label {299}
\\
\ga_{\th} & = & i \, \ee  \, + \, \th
\, \GG  + \, \tb \, \uu \, + \, \th \tb \, [ -i \ttt ]
\ \ \ .
\nonumber
\end{eqnarray}
{}From these equations
and the definition
$\gam = D \, {\rm ln} \, \ro_{z\zb}$, we conclude that
\begin{equation}
\ee \, = \, \ro_1 \, / \, \ro_0
\ \ \ \ , \ \ \ \
\GG \, = \, \pa \, {\rm ln} \, \ro_0
\ \ \ \ , \ \ \ \
\uu \, = \, i \, ( \ro_0 )^{-1} \, [ \, \ro_2 \, - \, \frac{i}{2} \,
\, \ro_1 \rob_1 \, / \, \ro_0 \, ]
\ \ \ \ , \ \ \ \
\ttt \, = \, \pa \, \eeb
\  .
\end{equation}
By virtue of eq.(\ref{666}), the space-time
field $\uu$ transforms homogeneously
under superconformal changes of coordinates and thereby
we can choose it to
vanish in the WZ-gauge:
according to the last set of equations, this simply
amounts to a redefinition of $\ro_2$ in terms of
the other components of the supermetric $\ro_{z\zb}$.
Thus, we are left with
$\GG \equiv \GG_z$ (which represents a non-holomorphic
affine connection) and its fermionic partner $\ee \equiv \ee_{\th}$
as well as the complex conjugate variables.

{}From the defining relation $\RR_{z \th} =
\frac{1}{2} \,
[ \pa \gam -
\frac{1}{2} \,
\gam (D\gam )]$ and the expansion
\begin{equation}
\RR_{z\th} \ = \ \frac{i}{2} \, \cc \, + \, \th
\, [ \frac{1}{2} \, \rr ] \, + \, \tb \,
[ \frac{1}{2} \, \uuu ] \, + \, \th \tb \,
[ -\frac{i}{2} \, \kk ]
\ \ \ ,
\end{equation}
it follows that
\begin{eqnarray}
\cc & = & (\pa - \frac{1}{2} \, \GG ) \, \ee
\ \equiv \ \chi (\GG)
\nonumber
\\
\rr & = & ( \pa \GG - \frac{1}{2} \, \GG ^2 )
\, - \, \frac{1}{2} \, \ee \  \chi (\GG)
\ \equiv \ r (\GG )
\, - \, \frac{1}{2} \, \ee \  \chi (\GG)
\ \equiv \ t (\GG )
\ \ \ .
\label{302}
\end{eqnarray}
Here, we introduced the functions $\chi ( \cdot ),
r( \cdot )$ and $t(\cdot )$ which may be viewed as `field
strengths' and which render the formulae below compact and transparent.

The supercovariant derivatives in superspace project down to
similar derivatives in space-time:
\begin{eqnarray}
(\nabla_{\ga} ^2 \HZB ) \! \! \mid  & = &
(\pa + \GG ) \, \mu
\, + \, \frac{1}{2} \, \ee \, \alpha  \ \equiv \ {\cal D} \mu
\nonumber
\\
(\nabla_{\ga} ^3 \HZB ) \! \! \mid  & = &  i \left[
(\pa + \frac{1}{2} \, \GG ) \, \alpha
+ (\, \frac{1}{2} \, \ee \, \pa \, + \, (\pa \ee ) \, ) \, \mu \right]
\ \equiv \ i \, {\cal D} \alpha
\ \ \ .
\label{303}
\end{eqnarray}
These expressions represent the `field strengths' of $\mu$ and $\alpha$
and they
appear for instance in the $\th$-component of $G$ as
defined by
eq.(\ref{37}):
\begin{equation}
\label{304}
(DG ) \! \! \mid \ = \ \GG \; - \; \frac{1}{1- \mu \bar{\mu}} \, \left[ \,
\bar{\cal D} \bar{\mu} \, - \, \bar{\mu}
\, {\cal D} \mu \, \right]
\ \equiv \ g
\ \ \ .
\end{equation}

The symmetry transformations of the basic space-time fields
follow from the superspace variations (\ref{18}),
\begin{eqnarray}
s \mu & = &  [ \, \pab \ - \ \mu \, \pa  \ + \
( \pa \mu ) \, ] \, c \ + \  \frac{1}{2} \, \alpha \, \epsilon
\nonumber
\\
s \alpha & = &
[ \, \pab \ - \ \mu \, \pa  \ + \ \frac{1}{2} \,
( \pa \mu ) \, ] \, \epsilon \ + \ c \,  \pa \alpha  \ - \ \frac{1}{2} \,
\alpha \, \pa c
\label{41}
\\
sc & = & - c \,  \pa c  \ + \  \frac{1}{4} \, \epsilon \, \epsilon
\nonumber
\\
s \epsilon & = & - c \,  \pa \epsilon  \ - \  \frac{1}{2} \, \epsilon
\, \pa c
\ \ \ ,
\nonumber
\end{eqnarray}
where we have the following assignments:
\begin{table}[h]\centering
\begin{tabular}{ || l || c | c | c | c || }  \hline\hline
Field   &  $\mu$   & $\alpha$   & $c$   &  $\epsilon$    \\  \hline
Grassmann parity  &  0  & 1   & 0   &  1    \\    \hline
Ghost number       &  0  & 0   & 1   &  1    \\    \hline\hline
\end{tabular}
\end{table}

As to the action of the $s$-operator, we adhere to the conventions
specified in the footnote in section 3. The $s$-operator
defined by eqs.(\ref{41}) is then nilpotent by construction.

Substitution of the $\th$-expansions into our previous superspace results
yields
\begin{eqnarray}
{\cal A}_D & = &
- \frac{K}{2}
\int_{{\bf \Sigma}} d^2z \
\left\{ \, c \, \left[ \; [ \pa^3 + 2r \pa + (\pa r) ] \, \mu\,
+ \, [ \frac{3}{2} \chi \pa + \frac{1}{2} (\pa \chi ) ] \, \alpha \; \right]
\right.
\nonumber
\\
& & \left.
\
\ \ \ \ \ \ \ \ \ \ \ \
\ \ \ \ \ \ \ \ \ \ \ \ \
- \ \epsilon \, \left[ \; [ \pa^2 + \frac{1}{2} r ] \, \alpha
\, + \, [ \frac{3}{2} \chi \pa +  (\pa \chi ) ] \,  \mu \; \right]
\right\}
\nonumber
\\
\Gamma_{II} & = &
- \frac{K}{2} \,
\int_{{\bf \Sigma}} d^2z \
\left\{  \, \mu  \, ( r - \rr ) \, - \, \alpha \, ( \chi - \cc ) \, \right\}
\ + \ {\rm c.c.}
\label{310}
\\
\Gamma_{III} & = &
\frac{K}{2} \,
\int_{{\bf \Sigma}} d^2z \ \,
\frac{1}{1 - \mu \bar{\mu}} \ \left\{ \,
({\cal D} \mu ) \, ( \bar{\cal D} \bar{\mu} )
\; - \; \frac{1}{2} \,
\bar{\mu} \, ( {\cal D} \mu )^2
\; - \; \frac{1}{2} \,
\mu \, ( \bar{\cal D} \bar{\mu} )^2
 \, \right\}
\ \ \ .
\nonumber
\end{eqnarray}
Using the $s$-transformations of $\mu$ and $\alpha$,
the $s$-invariance
of $\GG$ and $\ee$ as well as some
integration by parts, we find that
\begin{eqnarray}
s\Gamma_{III} & = &
\frac{K}{2} \,
\int_{{\bf \Sigma}} d^2z \
\left\{  \, (s\mu ) \,
[r(g ) - r (\GG ) ]  \, - \,
(s  \alpha ) \,
[\chi(g ) - \chi (\GG ) ]
\, \right\}
\; + \; {\rm c.c.}
\label{330}
\\
& = &
\frac{K}{2} \,
\int_{{\bf \Sigma}} d^2z \
\left\{  c \left[ \, -
[ \pab - \mu \pa - 2 (\pa \mu ) ] \,
[r(g ) - r (\GG ) ]
\, - \,
[ \frac{1}{2}  \alpha \pa  + \frac{3}{2} (\pa \alpha) ]
\, [\chi(g ) - \chi (\GG ) ] \,
\right]  \right.
\nonumber
\\
& & \left.
\ \ \ \ \ \ \ \ \ \ \ \ \ \
- \; \epsilon \,
\left[ \; \frac{1}{2} \,
[r(g ) - r (\GG ) ]
\, \alpha
\, - \,
[ \pab - \mu \pa - \frac{3}{2} (\pa \mu ) ]
\, [\chi(g ) - \chi (\GG ) ]
\; \right]
\, \right\}
\; + \; {\rm c.c.}
\ ,
\nonumber
\end{eqnarray}
where $g$ was defined in eq.(\ref{304}) and where
$\chi ( \cdot )$ and $r( \cdot )$ refer to the notation introduced
in eqs.(\ref{302}).

In the WZ-gauge,
the expression (\ref{39}) takes the form
\begin{eqnarray}
{\rm expression} \, (\ref{39})
& = &
- \int_{{\bf \Sigma}} d^2z \,
\left\{  \, c \, \left[  \, \pa^3 \mu \, - \,
[ \pab - \mu \pa - 2 (\pa \mu ) ] \, t(g) \, -  \,
[ \frac{1}{2} \, \alpha \pa + \frac{3}{2} (\pa \alpha ) ] \, \chi (g) \,
\right]  \right.
\nonumber
\\
& & \left.
\ \ \ \ \ \
- \; \epsilon \,
\left[ \, [\pa ^2  + \frac{1}{2} t(g) ] \, \alpha
\, - \,
[ \pab - \mu \pa - \frac{3}{2} (\pa \mu ) ] \, \chi (g) \, \right]
\, \right\}
\; + \; {\rm c.c.}
\ \ \ ,
\label{331}
\end{eqnarray}
where $t( \cdot )$ was defined in eq.(\ref{302}). Furthermore, equation
(\ref{102}) becomes
\begin{eqnarray}
s \Gamma_{II} & = &      {\cal A}_D  \;  + \;
\frac{K}{2}
\int_{{\bf \Sigma}} d^2z \,
\left\{  \, c \, \left[  \, \pa^3 \mu \, -\,
[ \pab - \mu \pa - 2 (\pa \mu ) ] \, \rr
\, - \,
[ \frac{1}{2}  \alpha \pa  + \frac{3}{2} (\pa \alpha ) ] \, \cc \,
\right]  \right.
\label{332}
\\
& & \left.
\ \ \ \ \ \ \ \ \ \
\ \ \ \ \ \  \ \ \ \ \ \
\ \ \ \ \ \  \ \ \ \ \ \
- \, \epsilon\,
\left[ \, [\pa ^2  + \frac{1}{2} \rr ] \, \alpha
\, -  \,
[ \pab - \mu \pa - \frac{3}{2} (\pa \mu ) ]  \, \cc
\, \right]
\, \right\}
\; + \; {\rm c.c.}
\ \ .
\nonumber
\end{eqnarray}
Since the expression
(\ref{331}) represents the WZ-gauge version of the integral on the
r.h.s. of eq.(\ref{30}), it is readily seen from the latter formula
and eqs.(\ref{330})(\ref{332}) that the result (\ref{500}) holds
in the WZ-gauge. In particular, it encompasses the results of the
bosonic theory \cite{klt}.

\section{Conclusion}

By virtue of a technical {\em tour de force}, we have constructed the
local counterterm relating the super Weyl anomaly with
the chirally split superdiffeomorphism anomaly
(as defined on a compact SRS).
This result explicitly proves the equivalence of
both anomalies which expressions have been known
for some time
and discussed in the literature.
As by-products of our construction, we obtained the non-chirally split
superdiffeomorphism
anomaly and the super Verlinde action which are of independent
interest. The combination of the second and third counterterm
can be used to derive a holomorphic factorization theorem for arbitrary
central charge
along the lines of reference \cite{kls}.
This derivation
involves a discussion of
renormalized determinants and the index theorem for families and is to be
discussed separately \cite{ip}.

\vskip 1.5truecm

{\bf \Large{Acknowledgements}}

\vspace {5mm}

J.-P.A. is grateful to the Theory Division of CERN for their
hospitality and financial support during a stay in autumn
1991 where part of this work has been done.
F.G. wishes to thank the Theory Group of Bordeaux
for their hospitality at an early stage of this work.

\newpage

\appendix

\section{Intermediate coordinates}

In the following, we recall the basic relations for the
so-called intermediate or tilde coordinates \cite{dg}
and we derive some useful equations which find a direct application in the
main text.

The tilde coordinates are introduced in analogy
to the capital coordinates by
considering smooth transformations,
\[
\left( z, \zb , \th , \tb \right)
\longrightarrow
\left( \zt, \zbt , \tht , \tbt \right)  \equiv
\left( Z, \zb , \TH , \tb \right)
\longrightarrow
\left( Z, \ZB , \TH , \TB \right)
\ \ \ .
\]
They are {\em not} related to each other by
complex conjugation. For $\HT = 0$ (and c.c.),
we have the relations
\begin{eqnarray}
\pat & = & \frac{1}{\LA} \left[ \pa - ( D \, {\rm ln }
\, \sqrt{\LA} ) D \right]
\nonumber   \\
\pabt & = & \pab - \HZB \pa - \HZ D
\nonumber   \\
\dt & = & \frac{1}{\sqrt{\LA}} \, D
\label{a1}
\\
\dbt & = & \dab - \HB \pa - \HOB D
\ \ \ ,
\nonumber
\end{eqnarray}
by virtue of which the IFEQ's (\ref{15}) take the compact form
\begin{equation}
\label{a1b}
\dbt \, {\rm ln} \, \LA \ = \ \pa \HB
\ \ \ \ \ \ \ , \ \ \ \ \ \
\pabt \, {\rm ln} \, \LA \ = \ \pa \HZB
\ \ \ .
\end{equation}
Quite generally, the tilde and capital coordinates are
related by \cite{dg}
\begin{eqnarray}
\pat & = & \pa_Z + (\KZB \LB ) \, \pa_{\ZB} +
(\KZB \BT + \KZ \sqrt{\LB}) \, D_{\TB}
\nonumber   \\
\pabt & = &  \LB \, \pa_{\ZB} + \BT \, D_{\TB}
\nonumber   \\
\dt & = &  D_{\TH} + (\KB \LB ) \, \pa_{\ZB}
+ (\KB \BT + \KOB \sqrt{\LB}) \, D_{\TB}
\label{a1a}
\\
\dbt & = &
(\KT \BT + \KO \sqrt{\LB}) \, D_{\TB}
+ (\KT \LB ) \, \pa_{\ZB}
\ \ \ ,
\nonumber
\end{eqnarray}
or
\begin{eqnarray}
\pa_Z & = &
\pat - \HKZB \pabt - \HKZ \dbt
\nonumber   \\
\pa_{\ZB} & = &
\frac{1}{\bar{L}} \, \left[ \pabt -
\frac{\bar{T}}{\sqrt{\bar{L}} \KO} \,
(\dbt - \KT \pabt ) \right]
\nonumber   \\
D_{\TH} & = &
\dt - \HKB \pabt - \HKOB \dbt
\label{a2}
\\
D_{\TB} & = &
\frac{1}{\sqrt{\bar{L}} \KO} \, \left[
\dbt - \KT \pabt  \right]
\ \ \ ,
\nonumber
\end{eqnarray}
with
\begin{eqnarray}
\HKZ & \equiv &
(\KO )^{-1} \, \KZ
\ \ \ \ \ \ \ , \ \ \ \ \ \
\HKZB \ \equiv \
\KZB - \HKZ \, \KT
\nonumber   \\
\HKOB & \equiv &
(\KO )^{-1} \, \KOB
\ \ \ \ \ \ \ , \ \ \ \ \ \
\HKB \ \equiv \
\KB - \HKOB \, \KT
\ \ \ .
\label{a3}
\end{eqnarray}
For $\HT = 0$ (and c.c.), we have the explicit expressions
\begin{eqnarray}
\bar{L} & = & \bar{\LA} \; A
\ \ \ \
\ \ \ \ \ \ \ \ \ \
\ \ \ \ \ , \ \ \ \ \ \
\KT \ = \ - A^{-1} \; ( \HB H_z ^{\ \zb} + \HOB H_{\th}^{\ \zb})
\nonumber  \\
\KB & = & \LA ^{-1/2} \, A^{-1} \; H_{\th} ^{\ \zb}
\ \ \ \ \ \ , \ \ \ \ \ \
\KZB \ = \ \LA^ {-3/2} \, A^{-1} \; ( H_z ^{\ \zb} \sqrt{\LA}
+ H_{\th} ^{\ \zb} \tau )
\label{a4}
\\
\bar{T} & = & A \;  \bar{\tau} - ( \HZB H_z^{\ \tb} + \HZ
H_{\th} ^{ \ \tb} ) \sqrt{\bar{\LA}}
\nonumber   \\
\KZ & = &
\LA^{-3/2} \; A^{-1/2} \, \left[
( H_z^{\ \tb} \sqrt{\LA} - H_{\th} ^{\ \tb} \tau )
+ A^{-1} \,
( H_z^{\ \zb} \sqrt{\LA} + H_{\th} ^{\ \zb} \tau )
( \HZB H_z^{\ \tb}  + \HZ H_{\th} ^{\ \tb}  )  \right]
\nonumber   \\
\KOB & = &
\LA^{-1/2} \; A^{-1/2} \, \left[
H_{\th} ^{\ \tb} + A^{-1} \, H_{\th} ^{\ \zb} \,
( \HZB H_z^{\ \tb}  + \HZ H_{\th} ^{\ \tb}  )  \right]
\nonumber   \\
\KO & = &
A^{-1/2} \, \left[
1 - ( \HOB H_{\th} ^{\ \tb} + \HB H_z ^{\ \tb} ) \, - \,
A^{-1} \,
(\HB H_z ^{\ \zb} + \HOB H_{\th} ^{\ \zb} )
( \HZB H_z^{\ \tb}  + \HZ H_{\th} ^{\ \tb}  )  \right]
\ \ \ ,
\nonumber
\end{eqnarray}
where
\begin{equation}
\label{157}
A \; \equiv \; 1 - \left(
\HZB H_z ^{\ \zb} + \HZ H_{\th}^{\ \zb} \right)
\ \ \  .
\end{equation}
The Jacobians for the transformations (\ref{a1}) and (\ref{a2})
are given by $\sqrt{\LA}$ and $\sqrt{\bar{L}} / \KO$, respectively.

{}From the structure relations
\begin{equation}
\label{a5}
\{ D_{\TH} , D_{\TH} \} \ = \ 2\, \pa_Z
\ \ \  ,  \ \ \
\{ D_{\TB} , D_{\TB} \} \ = \ 2\, \pa_{\ZB}
\ \ \   , \ \ \
{\rm all \ other \ graded \ commutators}\ = \ 0
\ \ ,
\end{equation}
one can derive useful relations between the
`$k$' and $\bar{L}, \bar{T}$, e.g.
\begin{eqnarray}
D_{\TB} \KT & = &
\frac{1}{\sqrt{\bar{L}} \KO} \, \left[
1 - (\KO )^2 \right]
\ \ \ \ \ \ , \ \ \ \ \ \
D_{\TB} \KO \ = \
-\, \frac{1}{2 \sqrt{\bar{L}}} \, \pabt \KT
\nonumber  \\
D_{\TB} \left(
\frac{1}{\sqrt{\bar{L}}}  \right) & = &
- \, \frac{1}{\bar{L}} \left[
\frac{\bar{T}}{\sqrt{\bar{L}}}  +
\frac{\pabt \KT }{2  \KO} \, \right]
\ \ \ \ \ \ , \ \ \ \ \ \
\dbt \left( \frac{\KT}{(\KO )^2} \right) \ = \
\frac{1}{(\KO )^2} \ \left[ 1 - (\KO )^2 \right]
\nonumber   \\
( \KO )^2 & = & 1 - \dbt \KT + \KT \pabt \KT
\, \ \ \ \ \ \ ,
\ \ \ \ \ \
\left( \dbt - \KT \pabt \right) ^2 \ = \ (\KO )^2 \, \pabt
\label{a6}
\\
D_{\TH} \HKB & = &  \HKZB + (\HKOB )^2
\ \ \ \ \ \
\ \ \ \ \ \
\; \ \ \ \ , \ \ \ \ \ \ \,
D_{\TH} \HKOB \ = \  \HKZ
\nonumber
\end{eqnarray}
and
\begin{eqnarray}
D_{\TB} (\KZB \LB ) - 2\, ( \KZB \BT + \KZ \sqrt{\LB})
& = &
\sqrt{\LB} \ \,
\frac{\pat \KT}{\KO}
\nonumber  \\
D_{\TB} (\KB \LB ) + 2\, ( \KB \BT + \KOB \sqrt{\LB})
& = &
- \, \sqrt{\LB} \ \,
\frac{\dt \KT}{\KO}
\ \ \ .
\label{a7}
\end{eqnarray}

The IFEQ's (\ref{a1b}) allow us to evaluate $D_{\TH} \, {\rm ln} \, \bar{\LA}$.
The resulting expression involves some lengthy terms proportional
to $\dab \,
{\rm ln} \, \bar{\LA}$ and $\pab \,
{\rm ln} \, \bar{\LA}$. However, these terms vanish by virtue of the relations
between the `$k$' and `$H$' and one is left with a differential
polynomial in the `$H$' (up to a factor $1/\sqrt{\LA}$):
\begin{equation}
\label{a9}
D_{\TH} \, {\rm ln} \, \bar{\LA}  =  \frac{1}{\sqrt{\LA}} \left\{
\left[ 1 +
( \sqrt{\LA} \HKOB ) \HOB +
( \sqrt{\LA} \HKB ) \HZ \right] \pab H_{\th} ^{\ \zb}
+  \left[
( \sqrt{\LA} \HKOB ) \HB +
( \sqrt{\LA} \HKB ) \HZB \right] \pab H_{z} ^{\ \zb}
\right\}
{}.
\end{equation}
The complex conjugate expression immediately follows from
eqs.(\ref{a2})(\ref{a1b}) and (\ref{a4}):
\begin{equation}
\label{a9a}
D_{\TB} \, {\rm ln} \, \LA \ = \
\frac{1}{\sqrt{\bar{\LA}}} \;  \left\{ \,
\frac{1}{\sqrt{A} \, \KO } \,
\left[  \, \pa \HB - \KT \, \pa \HZB \, \right] \, \right\}
\ .
\end{equation}

Since the `$k$' and the dependent `$H$'
involve derivatives of the basic variable
$\HB$, they transform in a
complicated, non-homogenous way under a superconformal
change of coordinates, eqs.(\ref{23a}).
Fortunately, the functionals of interest to us (like the super Verlinde
action (\ref{601})) involve
combinations of these variables which transform in a simple way:
\begin{eqnarray}
A^{\prime} & = & A \, \left[ \, 1 \, + \, (\dab \bar{w} ) \, \KT \, \right]
\ \   ,   \ \
(\KO )^{\prime} \ = \  \KO \, + \,  \frac{1}{2} \,
(\dab \bar{w} ) \, A^{-1/2} \, \KT \,
(1 - \HOB H_{\th} ^{\ \tb} - \HB H_z ^{\ \tb} )
\nonumber
\\
(\sqrt{\LA} \, \HKB )^{\prime} & = &
{\rm e} ^w \;
{\rm e} ^{- 2 \bar{w}} \;
(\sqrt{\LA} \, \HKB )
\ \ \ \ \ ,  \ \ \
(\KT )^{\prime} \ = \ {\rm e} ^{- \bar{w}} \; \KT
\label{a21}
\\
(\sqrt{\LA} \, \HKOB )^{\prime} & = &
{\rm e} ^w \;
{\rm e} ^{- \bar{w}} \, \left[ \,
(\sqrt{\LA} \, \HKOB )  \, + \,
(\dab \bar{w} ) \,
(\sqrt{\LA} \, \HKB ) \, \right]
\ \ \ .
\nonumber
\end{eqnarray}

\section{Relating small and capital coordinates}

As in the main text, we consider the case $\HT = 0$ (and c.c.).
The particular relation between the small and capital coordinates
as expressed by eqs.(\ref{a2}) implies that a functional written in terms
of the capital coordinates takes a particular form when expressed in terms
of the small coordinates: generically, one obtains an integral of the
form
\begin{equation}
\label{d1}
\int d^4 z \ \frac{1}{(\KO )^2} \ \left[ \, {\cal P} (\HB ) \, + \,
\KT \, {\cal P} (\HZB ) \, \right]
\ \ \ ,
\end{equation}
where ${\cal P} (\HB )$ denotes the action of a differential
operator ${\cal P}$ on $\HB$. For instance, for the super
$bc$-system \cite{df},
the authors of reference \cite{dg} found that
\begin{equation}
\label{d2}
\int d^4 Z \ {\cal B}_{\TH Z} \, D_{\TB} {\cal C}^Z \ = \
- \int d^4 z \ \frac{1}{(\KO )^2} \ B_{\th z} \,
\left[ \, s\HB  \, - \,
\KT \, s \HZB \, \right]
\ \ \ ,
\end{equation}
where $s$ denotes the BRS operator and
\[
{\cal B}_{\TH Z} \, = \, \LA^{-3/2} \, B_{\th z }
\ \ \ \ \ \ , \ \ \ \ \ \
{\cal C}^Z \, = \, \LA \, C^z
\ \ \ .
\]
Further examples are provided by the derivations in the main body of the text
and will explicitly be given below.

By virtue of eq.(\ref{14}), we can express $\HZB$ in terms of the
independent variable $\HB$ as
$\HZB = \dbt \HB - 1/4 \, (D\HB)^2$ and use integration by parts
to transform
${\cal P}(\HZB )$ into an expression involving
${\cal P}(\HB )$.
Application of the relation between $\KO$ and $\KT$ given by
eqs.(\ref{a6}) then yields
\begin{equation}
\label{d3}
\int d^4 z \ \frac{1}{(\KO )^2} \ \left[ \, {\cal P} (\HB ) \, + \,
\KT \, {\cal P} (\HZB ) \, \right]
\ = \
\int d^4 z \  {\cal P} (\HB ) \  + \
\int d^4 z \ \frac{\KT}{(\KO )^2} \ [ \, ... \, ]
\ \ \ ,
\end{equation}
where the second term vanishes in the WZ-gauge.
E.g. the expression (\ref{d2}) can be cast into the form
\begin{equation}
\label{d4}
- \int d^4 z \  B_{\th z} \,   s\HB  \  - \
\int d^4 z \ \frac{\KT}{(\KO )^2} \
\left[ \, \dbt B_{\th z}
- \frac{3}{2} (\pa \HB ) B_{\th z} \, \right]
\  ( s\HB )
\end{equation}
and, in the WZ-gauge, this quantity equals
$\int d^2 z \  \left\{ \,  b_{zz} \,   s\mu \,  + \,
\beta_{\th z} \,   s\alpha \,  \right\}$ where
we used the $\th$-expansion
$B_{\th z} = i \beta_{\th z} + \th \, b_{zz}  +..$.

Let us now come to the examples encountered in the present paper.
By substituting eqs.(\ref{37}) and (\ref{40}) into (\ref{39}), we
end up with the following expression for the integral on the r.h.s.
of eq.(\ref{30}):
\begin{eqnarray}
\label{d5}
& &
K
\int_{{\bf S\Sigma}}
d^4 z
\frac{1}{(\KO)^2}  \  \left( \ C^z
\left\{
\, \frac{1}{2} \, \pa^2 D \HB + \left[ \dab - \HB \pa +
\frac{1}{2} (D\HB) D -
\frac{3}{2} (\pa \HB) \right] \RR \,
\right\}  \right.
\\
 &   &   \left.
\ \ \ \ \ \
\ \ \ \ \ \ \ \
+ \ \KT    \  C^z
\left\{
\,  \frac{1}{2} \,  \pa^2 D \HZB - \left[ \pab - \HZB \pa -
\frac{1}{2} (D\HZB) D -
\frac{3}{2} (\pa \HZB) \right] \RR \,
\right\}  \right)
\; + \; {\rm c.c.} \; + \; ...
\nonumber
\\
& &
\ \ \ \ \ \ \ \
= \ s \, \Gamma_X \ + \ [ \, {\cal A}_X \, + \, {\rm c.c.} \, ] \ + \ ...
\ \ \ .
\nonumber
\end{eqnarray}
Here,
\begin{eqnarray}
\label{d6}
{\cal A}_{X} [ C^z , \HB ]
& \equiv & \frac{K}{2}  \,
\int_{{\bf S\Sigma}} d^4z \
\frac{1}{(\KO )^2} \ C^z \, \left\{
[ \pa^2 D + 3 {\cal R} \pa + (D {\cal R} ) D
+ 2 (\pa {\cal R} ) ] \, \HB  \right.
\\
& & \left. \ \ \ \ \ \ \ \ \ \ \ \ \ \ \ \ \ \ \ \ \
\ \ + \ \KT \,
[ \pa^2 D + 3 {\cal R} \pa + (D {\cal R} ) D
+ 2 (\pa {\cal R} ) ] \,
\HZB \, \right\}
\nonumber
\end{eqnarray}
and
\begin{equation}
\label{d7}
\Gamma_{X} \ \equiv \ K \, \int_{{\bf S\Sigma}} d^4z \,
\frac{1}{(\KO )^2} \ \left[
\HB - \KT \HZB \right]
\, ( {\cal R} - \RR ) \ + \ {\rm c.c.}
\ \ \ .
\end{equation}
Clearly, the expressions (\ref{d5})-(\ref{d7}) are all of the form (\ref{d1}).
By the procedure outlined above, we find
\begin{eqnarray}
\label{d8}
{\cal A}_{X} & = &
{\cal A}_{D} \ + \
{\cal A}_{Q}
\\
{\cal A}_{Q} & \equiv &
\frac{K}{2}  \,
\int_{{\bf S\Sigma}} d^4z \
\frac{\KT}{(\KO )^2} \, \left\{  (s \HB )
[ \pa^2 D + 3 {\cal R} \pa + (D {\cal R} ) D
+ 2 (\pa {\cal R} ) ] \, \HB  \right.
\nonumber
\\
& & \left. \ \ \ \ \ \ \ \ \ \ \ \ \ \ \ \ \ \ \ \ \
\ \ \ \ \ \
\ \ - \ 2 C^z \HB
[ (\pa \HB )(\pa {\cal R}) - (D\HB )(\pa D {\cal R} ) ]
\right\}
\ \ \ ,
\nonumber
\end{eqnarray}
(where ${\cal A}_{D}$ is the chirally split superdiffeomorphism anomaly,
eq.(\ref{20})) and
\begin{eqnarray}
\Gamma_{X} & = &
- \, \Gamma_{II} \ - \
\Gamma_{P}
\label{d9}
\\
\Gamma_{P}  & \equiv  &
 K \, \int_{{\bf S\Sigma}} d^4z \
\frac{\KT}{(\KO )^2} \, \left\{  \dbt ({\cal R} - \RR ) \HB - \frac{1}{2}
({\cal R} - \RR )
[  \HB (\pa \HB ) + \frac{1}{2} (D\HB )^2  ]  \right\}
\ + \ {\rm c.c.}
\ ,
\nonumber
\end{eqnarray}
where $\Gamma_{II}$ represents the
counterterm introduced in eq.(\ref{101a}). Furthermore,
the integral in eq.(\ref{d5}) can be rewritten as
\begin{eqnarray}
& K &
\int_{{\bf S\Sigma}}
d^4 z
\  C^z \left\{
\, \frac{1}{2} \, \pa^2 D \HB + \left[ \dab - \HB \pa +
\frac{1}{2} (D\HB) D -
\frac{3}{2} (\pa \HB) \right] \RR \,
\right\}
\\
+  & K &
\int_{{\bf S\Sigma}}
d^4 z  \
\frac{\KT}{(\KO)^2}
\;  (s\HB ) \,
\left\{
\, \frac{1}{2} \, \pa^2 D \HB + \left[ \dab - \HB \pa +
\frac{1}{2} (D\HB) D -
\frac{3}{2} (\pa \HB) \right] \RR \,
\right\}
\ \ \ ,
\nonumber
\end{eqnarray}
where the first contribution is the one specified in eq.(\ref{42}).

\section{Technical details}

In this appendix, we provide some computational details
for the derivations of section 5.

{}From eqs.(\ref{35}), (\ref{a5}), (\ref{33}), (\ref{a1a}), (\ref{17}),
(\ref{a1}) and integration by parts, it follows that the expression
(\ref{32}) can be rewritten
(up to the `c.c.' contribution) as
\begin{eqnarray}
\label{b1}
& &
\int_{{\bf S\Sigma}}
d^4Z \ {\cal C}^Z D_{\TB}
\left\{ \left[ \pat +
[ D_{\TB} (\KZB \LB ) - 2\, ( \KZB \BT + \KZ \sqrt{\LB} ) ]
\, D_{\TB} \, \right] \GAM \right.
\\
& & \left.  \ \ \ \ \ \ \ \ \ \
- \frac{1}{2} \, \GAM \left[ \dt - [
D_{\TB} (\KB \LB ) + 2\, ( \KB \BT + \KOB \sqrt{\LB})
] \, D_{\TB} \, \right]  \GAM
- \frac{1}{2} \, (\KB \LB ) \,
(D_{\TB} \GAM )^2 \right\}
 \ .
\nonumber
\end{eqnarray}
Next, substitution of (\ref{a7}) and (\ref{a1a}) into the
previous integral yields the expression (\ref{34})
(up to the c.c.).

The passage between eqs.(\ref{35}) and (\ref{36}) gives rise to
complicated expressions which are proportional to
$\ga_{\tb}, \dab \ga_{\tb}, \dab \, {\rm ln} \, \bar{\LA}$
and
$\pab \, {\rm ln} \, \bar{\LA}$, respectively. Each
of these expressions can be shown
to vanish by virtue of the relations between $\LB, \BT$ and the `$k$'
which follow from the structure equations.

The derivation of eq.(\ref{39}) from (\ref{34}) relies
on the formulae (\ref{a6}) and on the
commutation relations between $D_{\TB}$
and the tilde derivatives.

\end{document}